\documentclass[a4paper,12pt,eqno]{article}
\usepackage{theorem}
\usepackage{latexsym,amssymb,amsfonts,amsmath}
\usepackage{graphicx}
\makeatletter
\@addtoreset{equation}{section}

\makeatother

\newtheorem{thm}{Theorem}[section]
\newtheorem{lem}[thm]{Lemma}
\newtheorem{cor}[thm]{Corollary}
\newtheorem{pro}[thm]{Proposition}

\theoremheaderfont{\scshape}
\newcommand{\RM}{\mathbb{R}}
\newcommand{\ZM}{\mathbb{Z}}

\newcommand{\CM}{\mathbb{C}}

\newcommand{\ket}[1]{|#1\rangle}

\title{{\Large {\bf The non-uniform stationary measure for discrete-time quantum walks in one dimension}}}
\author{
{\small Norio Konno$^{(1)}$, $\quad$ Masato Takei$^{(2)}$}\\
{\scriptsize Department of Applied Mathematics, 
Faculty of Engineering, 
Yokohama National University}\\
{\scriptsize Hodogaya, Yokohama 240-8501, Japan}\\
{\scriptsize (1) e-mail: konno@ynu.ac.jp, $\quad$ 
(2) e-mail: takei@ynu.ac.jp}\\
}
\date{\empty }
\begin{document}
\maketitle

\par\noindent
\begin{small}
\par\noindent
{\bf Abstract}. We consider stationary measures of the one-dimensional discrete-time quantum walks (QWs) with two chiralities, which is defined by a 2 $\times$ 2 unitary matrix $U$. In our previous paper \cite{Konno2014}, we proved that any uniform measure becomes the stationary measure of the QW by solving the corresponding eigenvalue problem. This paper reports that non-uniform measures are also stationary measures of the QW except $U$ is diagonal. For diagonal matrices, we show that any stationary measure is uniform. Moreover, we prove that any uniform measure becomes a stationary measure for more general QWs not by solving the eigenvalue problem but by a simple argument. 
  
\footnote[0]{
{\it Abbr. title:} Non-uniform measure for quantum walks
}
\end{small}

\setcounter{equation}{0}
\section{Introduction \label{intro}}
The quantum walk (QW) is a quantum version of the classical random walk. QWs have been largely investigated for the last decade. The review and books on QWs are, for example, Kempe \cite{Kempe2003}, Kendon \cite{Kendon2007}, Venegas-Andraca \cite{VAndraca2008, Venegas2013}, Konno \cite{Konno2008b}, Cantero et al. \cite{CanteroEtAl2013}, Manouchehri and Wang \cite{MW2013}, Portugal \cite{P2013}. Let $\ZM$ be the set of integers. The present paper focuses on the discrete-time QW with two chiralities on $\ZM$, which was first intensively studied by Ambainis et al. \cite{AmbainisEtAl2001}. The property of the stationary measure of the Markov chain has been extensively and deeply investigated. However, the corresponding result for QW is almost not known. This is a motivation of this paper. 

From now on, we briefly review our previous related results. Let $\mbox{\boldmath{U}}(n)$ denote the set of $n \times n$ unitary matrices. In our sequential works, we have considered the stationary measure of discrete-time space-inhomogeneous QWs on $\ZM$ with two chiralities determined by a sequence of $\mbox{\boldmath{U}}(2)$, that is, $\{ U_x \in \mbox{\boldmath{U}}(2) : x \in \ZM \}$, where
\begin{align*}
U_x =
\begin{bmatrix}
a_x & b_x \\
c_x & d_x
\end{bmatrix},
\end{align*}
with $a_x, b_x, c_x, d_x \in \mathbb C$ and $\CM$ is the set of complex numbers. We call $U_x$ quantum coin at position $x \in \ZM$. In particular, we investigated the following four types (i) $\sim$ (iv) of discrete-time QWs on $\ZM$ by solving the corresponding eigenvalue problem, whose quantum coins are given by
\begin{align*}
U_x = 
\left\{ 
\begin{array}{cc}
U_+ & (x \ge 1), \\
U_0 & (x =0), \\
U_- & (x \le -1). 
\end{array}
\right.
\end{align*}

(i) Konno et al. \cite{KLS2013} gave a stationary measure of the QW whose quantum coin is given by 
\begin{align*}
U_0 = H (\sigma) = \frac{1}{\sqrt{2}} 
\begin{bmatrix}
1 & e^{i \sigma} \\
e^{-i \sigma} & -1
\end{bmatrix},
\qquad U_+ = U_- = H,
\end{align*}
where $\sigma \in [0,2\pi)$, and $H= H (0)$ is the Hadamard matrix, that is, 
\begin{align*}
H = \frac{1}{\sqrt{2}} 
\begin{bmatrix}
1 & 1 \\
1 & -1
\end{bmatrix}.
\end{align*}
Thus, if $\sigma = 0$, then this model becomes space-homogeneous and is equivalent to the {\it Hadamard walk}.

(ii) Endo and Konno \cite{EK2013} had a stationary measure of the QW given by
\begin{align*}
U_0 = \omega H, \qquad U_+ = U_- = H, 
\end{align*}
where $\omega = \exp(2 \pi i \phi)$ with $\phi \in [0,1)$. The model was introduced and studied by W\'ojcik et al. \cite{WojcikEtAl2004} and becomes the Hadamard walk if we take $\phi =0$.

(iii) Endo et al. \cite{EndoEtAl2014a} got a stationary measure of the QW determined by
\begin{align*}
U_0 = U (\theta), 
\qquad U_+ = U_- = H, 
\end{align*}
where
\begin{align*}
U (\theta) = 
\begin{bmatrix}
\cos \theta & \sin \theta \\
\sin \theta & - \cos \theta 
\end{bmatrix}, 
\end{align*}
with $\theta \in [0, 2 \pi)$. If $\theta = \pi/4$, then this model becomes the Hadamard walk.

(iv) Endo et al. \cite{EndoEtAl2014b} obtained a stationary measure of the QW given by 
\begin{align*}
U_0 = U (0), \quad U_+ = H (\sigma_+), \quad U_- = H (\sigma_-), 
\end{align*}
where $\sigma_+, \> \sigma_- \in [0, 2 \pi)$.

To explain our results, we introduce some notations. Let ${\cal M}_s$ be the set of stationary measures of the QW (the precise definition is given in the next section). For any $c>0$, $\mu_{u}^{(c)}$ denotes the uniform measure with parameter $c$, i.e., 
\begin{align*}
\mu_{u}^{(c)} (x) = c \qquad (x \in \ZM). 
\end{align*}
Let ${\cal M}_{unif} = \{ \mu_{u}^{(c)} : c>0 \}$ be the set of uniform measures on $\ZM$. Let ${\cal M}_{exp}$ be the set of the measures $\mu$ having exponential decay with respect to the position, i.e., $\mu$ satisfies that there exist positive constants $C_+, C_0, C_-$, and $\gamma \in (0,1)$ such that 
\begin{align*}
\mu (x) = 
\left\{ 
\begin{array}{cc}
C_+ \gamma^{-|x|} & (x \ge 1), \\
C_0 & (x =0), \\
C_- \gamma^{-|x|} & (x \le -1). 
\end{array}
\right.
\end{align*}
As for the above four models (i) $\sim$ (iv) with space-inhomogeneous quantum coin, all the stationary measures we obtained have exponential decay with respect to the position: For each QW, we see that
\begin{align*}
{\cal M}_s \cap {\cal M}_{exp} \not= \emptyset.
\end{align*}
So the stationary measures are not the uniform measures on $\ZM$. However, in a special space-homogeneous case (Hadamard walk), the stationary measure we got becomes the uniform measure.

Under this background, our previous paper \cite{Konno2014} treated the space-homogeneous QW whose quantum coin $U_x = U \in \mbox{\boldmath{U}} (2)$ determined by
\begin{align}
U =
\begin{bmatrix}
a & b \\
c & d
\end{bmatrix}.
\label{ohuro}
\end{align}
Remark that the unitarity of $U$ implies that it is enough to consider three cases: $abcd \not=0, \> a=0,$ and $b=0$. Then we obtained the following result for any $U \in \mbox{\boldmath{U}} (2)$ by solving the corresponding eigenvalue problem: 
\begin{thm}
\label{biwako1}
For any $U \in \mbox{\boldmath{U}} (2)$, we have
\begin{align}
{\cal M}_{unif} \subseteq {\cal M}_s.
\end{align}
\end{thm}
This paper is the continued one of \cite{Konno2014} and reports that some non-uniform measures are stationary measures of the QW (defined by $U$ of Eq. \eqref{ohuro}) except for $b=0$ case. Furthermore, these non-uniform measures do not have exponential decay. Thus we obtain the following result: 
\begin{thm}
\label{biwako2}
For any $U \in \mbox{\boldmath{U}} (2)$ with $abcd \not=0$ or $a=0$, we see 
\begin{align*}
{\cal M}_s \setminus \left({\cal M}_{unif} \cup {\cal M}_{exp} \right) \not= \emptyset.
\end{align*}
\end{thm}
For $b=0$ case, we show that any stationary measure is uniform. Thus combining this with Theorem \ref{biwako1} gives
\begin{thm}
\label{biwako3}
For any $U \in \mbox{\boldmath{U}} (2)$ with $b=0$, we see 
\begin{align*}
{\cal M}_s = {\cal M}_{unif}. 
\end{align*}
\end{thm}
Moreover, we prove that every uniform measure becomes stationary, i.e., Theorem \ref{biwako1} holds for more general QWs not by solving the eigenvalue problem but by a simple argument.

We should note that for the corresponding classical random walk in which the walker moves one step to the left with probability $p$ and to the right with probability $q$ with $p+q=1 \> (p,q \in [0,1])$, the uniform measure $\mu_{u}^{(c)} \> (c>0)$ is a stationary measure. Moreover, $(q/p)^x \> (x \in \ZM)$ with $pq \not=0$ and $p \not= q$ is a non-uniform stationary measure.

The rest of the present paper is organized as follows. Section \ref{def} gives the detailed definition of our two-state space-homogeneous model on $\ZM$. In the previous paper \cite{Konno2014}, we proved that ${\cal M}_{unif} \subseteq {\cal M}_s$ (referred to Theorem \ref{biwako1} in this paper) by solving the corresponding eigenvalue problem. Sections \ref{abcdzero} and \ref{azero} present non-uniform and non-exponential decay stationary measures for $abcd \not=0$ and $a=0$ by a similar method in \cite{Konno2014} respectively. In Sect. \ref{bzero}, we prove ${\cal M}_{s} = {\cal M}_{unif}$ for $b=0$. Section \ref{unitary} gives a new proof of Theorem \ref{biwako1}. In Sect. \ref{sum}, we summarize our results.

\section{Definition of Two-State Model \label{def}}
The discrete-time QW is a quantum version of the classical random walk with additional degree of freedom called chirality. The chirality takes values left and right, and it means the direction of the motion of the walker. At each time step, if the walker has the left chirality, it moves one step to the left, and if it has the right chirality, it moves one step to the right. Let us define
\begin{align*}
\ket{L} = 
\begin{bmatrix}
1 \\
0  
\end{bmatrix},
\qquad
\ket{R} = 
\begin{bmatrix}
0 \\
1  
\end{bmatrix},
\end{align*}
where $L$ and $R$ refer to the left and right chirality states, respectively.  

The time evolution of the walk is determined by $U \in \mbox{\boldmath{U}}(2)$, where $\mbox{\boldmath{U}}(n)$ be the set of $n \times n$ unitary matrices and  
\begin{align*}
U =
\begin{bmatrix}
a & b \\
c & d
\end{bmatrix}.
\end{align*}
To define the dynamics of our model, we divide $U$ into two matrices:
\begin{eqnarray*}
P =
\begin{bmatrix}
a & b \\
0 & 0 
\end{bmatrix}, 
\quad
Q =
\begin{bmatrix}
0 & 0 \\
c & d 
\end{bmatrix},
\end{eqnarray*}
with $U =P+Q$. The important point is that $P$ (resp. $Q$) represents that the walker moves to the left (resp. right) at any position at each time step.

One of the typical class considered here is 
\begin{align}
U = U (\theta) = 
\begin{bmatrix}
\cos \theta & \sin \theta \\
\sin \theta & - \cos \theta 
\end{bmatrix}, 
\label{akina}
\end{align}
where $\theta \in [0, 2 \pi)$. When $\theta = \pi/4$, then this model becomes the Hadamard walk. Let $\Psi_n$ denote the amplitude at time $n$ of the QW on $\ZM$:  
\begin{align*}
\Psi_{n}
&= {}^T\![\cdots,\Psi_{n}^{L}(-1),\Psi_{n}^{R}(-1),\Psi_{n}^{L}(0),\Psi_{n}^{R}(0),\Psi_{n}^{L}(1),\Psi_{n}^{R}(1),\cdots ],
\\
&= {}^T\!\left[\cdots,\begin{bmatrix}
\Psi_{n}^{L}(-1)\\
\Psi_{n}^{R}(-1)\end{bmatrix},\begin{bmatrix}
\Psi_{n}^{L}(0)\\
\Psi_{n}^{R}(0)\end{bmatrix},\begin{bmatrix}
\Psi_{n}^{L}(1)\\
\Psi_{n}^{R}(1)\end{bmatrix},\cdots\right],
\end{align*}
where $T$ means the transposed operation. Then the time evolution of the walk is defined by 
\begin{align*}
\Psi_{n+1}(x)= P \Psi_{n} (x+1) +  Q \Psi_{n}(x-1),
\end{align*}
where $\Psi_{n}(x)$ denotes the amplitude at time $n$ and position $x$. That is 
\begin{align*}
\begin{bmatrix}
\Psi_{n+1}^{L}(x)\\
\Psi_{n+1}^{R}(x)
\end{bmatrix}
=
\begin{bmatrix}           
a \Psi_{n}^{L}(x+1)+b \Psi_{n}^{R}(x+1)\\
c \Psi_{n}^{L}(x-1)+d \Psi_{n}^{R}(x-1)
\end{bmatrix}.
\end{align*}
Now let
\begin{align*}
U^{(s)}=\begin{bmatrix}
\ddots&\vdots&\vdots&\vdots&\vdots&\vdots&\cdots \\
\cdots&O&P&O&O&O&\cdots\\
\cdots&Q&O&P&O&O&\cdots\\
\cdots&O&Q&O&P&O&\cdots\\
\cdots&O&O&Q&O&P&\cdots\\
\cdots&O&O&O&Q&O&\cdots\\
\cdots&\vdots&\vdots&\vdots&\vdots&\vdots&\ddots
\end{bmatrix}\;\;\;
\mbox{with} \;\;\;
O=\begin{bmatrix}
0&0\\
0&0
\end{bmatrix}.
\end{align*}
Then the state of the QW at time $n$ is given by
\begin{align}
\Psi_{n}=(U^{(s)})^{n}\Psi_{0},
\label{sankeien}
\end{align} 
for any $n\geq0$. Let $\mathbb{R}_{+}=[0,\infty)$. Here we introduce a map 
$\phi:(\mathbb{C}^{2})^{\mathbb{Z}}\rightarrow \mathbb{R}_{+}^{\mathbb{Z}}$
such that if
\begin{align*}
\Psi= {}^T\!\left[\cdots,\begin{bmatrix}
\Psi^{L}(-1)\\
\Psi^{R}(-1)\end{bmatrix},\begin{bmatrix}
\Psi^{L}(0)\\
\Psi^{R}(0)\end{bmatrix},\begin{bmatrix}
\Psi^{L}(1)\\
\Psi^{R}(1)\end{bmatrix},\cdots\right] \in(\mathbb{C}^{2})^{\mathbb{Z}},
\end{align*}
then 
\begin{align*}
\phi(\Psi) = {}^T\! 
\left[\ldots, 
|\Psi^{L}(-1)|^2 + |\Psi^{R}(-1)|^2, 
|\Psi^{L}(0)|^2 + |\Psi^{R}(0)|^2, 
|\Psi^{L}(1)|^2 + |\Psi^{R}(1)|^2, \ldots \right] 
\in \mathbb{R}_{+}^{\mathbb{Z}}.
\end{align*}
That is, for any $x \in \ZM$, 
\begin{align*}
\phi(\Psi) (x) = |\Psi^{L}(x)|^2 + |\Psi^{R}(x)|^2.
\end{align*}
Sometimes we identify $\phi(\Psi(x))$ with $\phi(\Psi) (x)$. Moreover we define the measure of the QW at position $x$ by
\begin{align*}
\mu(x)=\phi(\Psi(x)) \quad (x \in \ZM).
\end{align*}
Now we are ready to introduce the set of stationary measures:  
\begin{align*}
{\cal M}_{s} 
&= {\cal M}_s (U)
\\
&= \left\{ \mu \in\mathbb{R}_{+}^{\mathbb{Z}} \setminus \{ \boldsymbol{0} \} : \mbox{there exists} \; \Psi_{0} \; \mbox{such that} \; \phi((U^{(s)})^{n}\Psi_{0})=\mu \; (n \ge 0) \right\},
\end{align*}
where $\boldsymbol{0}$ is the zero vector. We call the element of ${\cal M}_{s}$ the stationary measure of the QW. Moreover we introduce the following set of measures:
\begin{align*}
{\cal M}_{n} 
&= {\cal M}_n (U)
\\
&= \left\{ \mu \in\mathbb{R}_{+}^{\mathbb{Z}} \setminus \{ \boldsymbol{0} \} : \mbox{there exists} \; \Psi_{0} \; \mbox{such that} \; \phi((U^{(s)})^{k}\Psi_{0})=\mu  \; (k =0,1, \ldots, n) \right\}.
\end{align*}
By definition, we see that  
\begin{align*}
{\cal M}_{1} \supseteq {\cal M}_{2} \supseteq  \cdots \supseteq {\cal M}_{n} \supseteq {\cal M}_{n+1} \supseteq \cdots, \quad {\cal M}_s = \bigcap_{n=1}^{\infty} {\cal M}_n.
\end{align*}

Next we consider the eigenvalue problem of the QW:
\begin{align}
U^{(s)} \Psi = \lambda \Psi \quad (\lambda \in \mathbb{C}).
\label{samui}
\end{align}
Remark that $|\lambda|=1$, since $U^{(s)}$ is unitary. We sometime write $\Psi=\Psi^{(\lambda)}$ in order to emphasize the dependence on eigenvalue $\lambda$. Then we see that $\phi (\Psi^{(\lambda)}) \in {\cal M}_s$. Moreover we introduce 
\begin{align*} 
{\cal W}^{(\lambda)} 
= \left\{ \Psi^{(\lambda)} \in \CM^{\ZM} \setminus \{ \boldsymbol{0} \} : U^{(s)} \Psi^{(\lambda)} = \lambda \Psi^{(\lambda)} \right\}.
\end{align*}
We see that Eq. \eqref{samui} is equivalent to 
\begin{align}
\lambda \Psi^{L}(x) 
&= a \Psi^{L}(x+1) + b \Psi^{R} (x+1),
\label{yokoyama}
\\
\lambda \Psi^{R}(x)
&= c \Psi^{L}(x-1) + d \Psi^{R}(x-1),
\label{taikan} 
\end{align}
for any $x \in \ZM$.

\section{Case $abcd \not=0$ \label{abcdzero}}
Let $\Psi(x)={}^T \> [\Psi^{L}(x),\> \Psi^{R}(x)] \>\> (x \in \mathbb{Z})$ be the amplitude of the model at position $x$. In this section we present non-uniform stationary measures for $abcd \not=0$ by using the generating functions of $\Psi^{j}(x)$ for $j=L, R$. Here we introduce the generating functions for $\Psi^{L}(x)$ and $\Psi^{R}(x)$:
\begin{align*}
f^{j}(z)=\sum_{x \in  \mathbb{Z}} \Psi^{j}(x)z^{x} \quad(j=L,R).
\end{align*}
From Eqs. (\ref{yokoyama}) and (\ref{taikan}), we have obtained the following lemma (see Lemma 5.1 in \cite{Konno2014}).
\begin{lem}
We assume that 
\begin{align}
U^{(s)} \Psi = \lambda \Psi \quad (\lambda \in \mathbb{C}).
\end{align}
Then we obtain
\begin{align*}
A f (z) = 
\begin{bmatrix}
0 \\
0
\end{bmatrix},
\end{align*}
where 
\begin{align*}
A = 
\begin{bmatrix}
\lambda - \dfrac{a}{z} & -\dfrac{b}{z} \\
- cz & \lambda - dz
\end{bmatrix}, 
\qquad 
f (z) = 
\begin{bmatrix}
f^{L} (z) \\ 
f^{R} (z)
\end{bmatrix}.
\end{align*}
\end{lem}
Note that we have
\begin{align*}
\det A = - \frac{d \lambda}{z} h(z),
\end{align*}
where
\begin{align*}
h(z) = z^{2}- \frac{1}{d} \left(\lambda + \frac{\triangle}{\lambda} \right) z + \frac{a}{d},
\end{align*}
with $\triangle = \det U = ad -bc.$ Let $\phi \in (0, \pi/2)$ satisfy $\cos \phi = |a|, \> \sin \phi = \sqrt{1-|a|^2}$. Put $\xi \in [0, 2 \pi)$ with $\triangle = e^{i \xi}$. For the following four $\lambda$'s, $h(z)$ has double roots.
\begin{align*}
\lambda_1 = e^{i(\phi+(\xi/2))}, \> \lambda_2 = e^{i(-\phi+(\xi/2))}, \> \lambda_3 =  - \lambda_1, \> \lambda_4 = - \lambda_2.
\end{align*}

Moreover Eqs. \eqref{yokoyama} and \eqref{taikan} imply that $\Psi^{L}(x)$ and $\Psi^{R}(x)$ satisfy the following same equation:
\begin{align}
a_{x+2} - \frac{1}{a} \left(\lambda + \frac{\triangle}{\lambda} \right) a_{x+1} + \frac{d}{a} a_x = 0,
\label{aida}
\end{align}
for any $x \in \mathbb{Z}$. We generalize the argument in pp.1114-1115 of \cite{Konno2014}. We put
\begin{align}
\Psi^{L}(x) = (A +xB) \gamma^x \quad  (x \in \mathbb{Z}),
\label{hayami}
\end{align}
where $A, B \in \mathbb{C}$ with $|A|+|B| \not=0$. Here $\gamma \in \mathbb{C}$ is one of the double roots of the following characteristic polynomial for difference equation \eqref{aida}:
\begin{align*}
x^2 - \frac{1}{a} \left(\lambda + \frac{\triangle}{\lambda} \right) x + \frac{d}{a} = 0,
\end{align*}
Then we have
\begin{align}
\gamma = \frac{\lambda + \triangle \overline{\lambda}}{2a}.
\label{makoto}
\end{align}
From Eqs. \eqref{yokoyama} and \eqref{hayami}, we have
\begin{align*}
\Psi^{R}(x) = \left\{(A+xB) \left( \dfrac{\lambda - \triangle \overline{\lambda}}{2} \right) - \lambda B \right\} \dfrac{\gamma^{x-1}}{b} \quad  (x \in \mathbb{Z}).
\end{align*}
Then we see that for any $x \in \mathbb{Z}$,
\begin{align*}
\Psi^{L}(x) = (A+xB) \gamma^x, \quad \Psi^{R}(x) = \left\{(A+xB) \left( \dfrac{\lambda - \triangle \overline{\lambda}}{2} \right) - \lambda B \right\} \dfrac{\gamma^{x-1}}{b}. 
\end{align*}
In fact, $\Psi^{L}(x)$ and $\Psi^{R}(x)$ satisfy Eq. \eqref{taikan}. Thus we obtain the following result.

\begin{pro}
\label{kokoneAB}
For the QW with $abcd \not=0$, we see that
\begin{align*}
\Psi(x) = 
\begin{bmatrix}
\Psi^{L}(x) \\
\Psi^{R}(x)
\end{bmatrix}
=
\begin{bmatrix}           
(A+xB) \gamma^x, \\
\left\{(A+xB) \left( \dfrac{\lambda - \triangle \overline{\lambda}}{2} \right) - \lambda B \right\} \dfrac{\gamma^{x-1}}{b}
\end{bmatrix}
\quad  (x \in \mathbb{Z})
\end{align*}
satisfies
\begin{align*}
U^{(s)} \Psi = \lambda \Psi.
\end{align*}
\end{pro}

If we take the above $\Psi$ as the initial state $\Psi_0$, then we have 
\begin{align*}
\Psi_n = (U^{(s)})^n \Psi_0 = \lambda^n \Psi_0.
\end{align*}
Therefore we obtain the measure $\mu_n$ at time $n$ as follows:
\begin{align}
\mu_n (x) 
&= |\Psi_n^{L}(x)|^2 + |\Psi_n^{R}(x)|^2 = |\lambda|^{2n} \left( |\Psi_0^{L}(x)|^2 + |\Psi_0^{R}(x)|^2 \right) 
\nonumber
\\
&= \left\{ |A+xB|^2 |\gamma|^2 + \left|(A+xB) \left( \dfrac{\lambda - \triangle \overline{\lambda}}{2} \right) - \lambda B \right|^2 \frac{1}{|b|^2} \right\} |\gamma|^{2(x-1)}.
\label{taiyou}
\end{align}
From now on we compute $|\gamma|$. Equation \eqref{makoto} implies 
\begin{align}
|\gamma|^2 = \frac{ 1 + \Re (\triangle \overline{\lambda}^2)}{2 |a|^2},
\label{akiba}
\end{align}
where $\Re (z)$ is the real part of $z \in \CM$. On the other hand, for any $\lambda_k \> (k=1,2,3,4)$, we get
\begin{align}
\Re (\triangle \overline{\lambda}^2) = \cos (2 \phi) = 2 |a|^2 -1.
\label{otaku}
\end{align}
Combining Eq. \eqref{akiba} with Eq. \eqref{otaku} gives $|\gamma|=1$. Furthermore, in a similar fashion, we have
\begin{align}
\frac{|\lambda - \triangle \overline{\lambda}|^2}{4|b|^2} 
= \frac{1 - \Re (\triangle \overline{\lambda}^2)}{2 |b|^2}
=\frac{2(1-|a|^2)}{2|b|^2} = 1.
\label{kozou}
\end{align}
From Eq. \eqref{taiyou}, $|\gamma|=1$, and Eq. \eqref{kozou}, we obtain the measure $\mu_n$ for any $n \ge 0$ and $x \in \mathbb{Z}$ as follows.
\begin{cor}
\begin{align}
\mu_n (x) =2 |A+xB|^2 - 2 x |B|^2 + \frac{|B|^2 - \Re \left( A \overline{B} (1 - \triangle \overline{\lambda}^2) \right)}{|b|^2},
\label{Tokyotribe}
\end{align}
where $\Re (z)$ is the real part of $z \in \CM$. 
\end{cor}
So let $\mu=\mu_n$. Then this $\mu$ becomes stationary and (generally) non-uniform and non-exponential decay measure of the QW defined by $U$ with $abcd \not=0$. Therefore we have
\begin{align*}
{\cal M}_s (U) \setminus \left( {\cal M}_{unif} \cup {\cal M}_{exp} \right) \not= \emptyset.
\end{align*}

Here we consider the case of the QW determined by $U=U(\theta)$ with $0 < \theta < \pi/2$. In this case, we have $\phi = \theta$. Moreover $\xi = \pi$, since $\triangle = \det U(\theta) = -1$. Let 
\begin{align*}
\gamma_k = \frac{\lambda_k + \triangle \overline{\lambda_k}}{2 \cos \theta} \quad (k=1,2,3,4).
\end{align*}
So we have
\begin{align*}
\gamma_1 = \gamma_2 = i, \quad \gamma_3 = \gamma_4 = -i.
\end{align*}
For $k=1,2,3,4$, we put
\begin{align*}
\Psi^{(k)} (x) = 
\begin{bmatrix}
\Psi^{(k,L)}(x) \\
\Psi^{(k,R)}(x)
\end{bmatrix}.
\end{align*}
Therefore
\begin{align}
\Psi^{(1)} (x) 
&= 
\begin{bmatrix}
(A+xB) i^x \\
\left\{ (-A + B) - xB -i \cot \theta B \right\} i^{x-1}
\end{bmatrix},
\nonumber
\\ 
\Psi^{(2)} (x) 
&= 
\begin{bmatrix}
(A+xB) i^x \\
\left\{ (A - B) + xB -i \cot \theta B \right\} i^{x-1}
\end{bmatrix},
\nonumber
\\
\Psi^{(3)} (x) 
&= 
\begin{bmatrix}
(A+xB) (-i)^x \\
\left\{ (A - B) + xB +i \cot \theta B \right\} (-i)^{x-1}
\end{bmatrix},
\nonumber
\\ 
\Psi^{(4)} (x) 
&= 
\begin{bmatrix}
(A+xB) (-i)^x \\
\left\{ (-A + B) - xB +i \cot \theta B \right\} (-i)^{x-1}
\end{bmatrix}.
\label{taihu}
\end{align}
Then $\Psi^{(k)} \in {\cal W}^{(\lambda_k)}.$ For the Hadamard walk ($\theta = \pi/4$), if we take $A, B \in \RM$, then we have
\begin{align}
\mu_n (x) = 2 \left\{ B^2 x^2 + B(2A-B) x + A^2 - AB + B^2 \right\}.
\label{Tokyotribe2}
\end{align}
In particular, when $A=0$, 
\begin{align}
\mu_n (x) =2 \left\{ \left(x - \frac{1}{2} \right)^2 + \frac{3}{4} \right\} B^2.
\end{align}

Equations (\ref{taihu}) with $B=0$ become the following given in \cite{Konno2014}:
\begin{align*}
\Psi^{(1)} (x) 
&= 
\begin{bmatrix}
A i^x \\
- A i^{x-1}
\end{bmatrix},
\quad 
\Psi^{(2)} (x) = 
\begin{bmatrix}
A i^x \\
A i^{x-1}
\end{bmatrix},
\nonumber
\\
\Psi^{(3)} (x) 
&= 
\begin{bmatrix}
A (-i)^x \\
A (-i)^{x-1}
\end{bmatrix},
\quad 
\Psi^{(4)} (x) = 
\begin{bmatrix}
A (-i)^x \\
-A (-i)^{x-1}
\end{bmatrix}.
\end{align*}
Then $\Psi^{(k)} \in {\cal W}^{(\lambda_k)}.$ Thus $\mu_n (x) = 2 |A|^2$ for any $n \ge 0$ and $x \in \ZM$. Thus $\mu_n (x) = 2 |A|^2$ for any $n \ge 0$ and $x \in \ZM$.

\section{Case $a=0$ \label{azero}}
In this case, $U$ can be expressed as 
\begin{align*}
U=
\begin{bmatrix}
0 & e^{i \eta}  \\
- \triangle e^{-i \eta} & 0
\end{bmatrix},
\end{align*}
where $\eta \in [0, 2 \pi)$ and $\triangle (= \det U) \in \mathbb{C}$ with $|\triangle|=1$. 

From Eqs. \eqref{yokoyama} and \eqref{taikan}, we get
\begin{align*}
\lambda \Psi^{L}(x) 
&=  e^{i \eta}  \Psi^{R} (x+1),
\\
\lambda \Psi^{R}(x)
&= - \triangle e^{-i \eta} \Psi^{L}(x-1).
\end{align*}
By these equations, we see that for any $x \in \ZM$,
\begin{align*}
\left( 1 + \frac{\triangle}{\lambda^2} \right) \Psi^{j}(x) = 0 \quad (j=L,R).
\end{align*}
From this, we get $\triangle = - \lambda^2$. Let $\lambda_{\pm} = \pm i \sqrt{\triangle},$ where the sign is chosen in a suitable way. As an initial state, we consider $\Psi^{(\pm)}$ corresponding to $\lambda_{\pm}$ as follows;
\begin{align}
\Psi^{(\pm)} = {}^T \left[ \ldots, 
\begin{bmatrix} \Psi^{(\pm,L)} (-2) \\ \Psi^{(\pm,R)} (-2) \end{bmatrix}, 
\begin{bmatrix} \Psi^{(\pm,L)} (-1) \\ \Psi^{(\pm,R)} (-1) \end{bmatrix}, 
\begin{bmatrix} \Psi^{(\pm,L)} (0)  \\ \Psi^{(\pm,R)} (0)  \end{bmatrix}, 
\begin{bmatrix} \Psi^{(\pm,L)} (1)  \\ \Psi^{(\pm,R)} (1)  \end{bmatrix}, 
\begin{bmatrix} \Psi^{(\pm,L)} (2)  \\ \Psi^{(\pm,R)} (2)  \end{bmatrix}, 
\ldots \right].
\label{huyumiHayaku}
\end{align}
Here for any $x \in \ZM$, 
\begin{align}
\Psi^{(\pm,L)} (2x) &= \alpha_{2x}, \>\> \Psi^{(\pm,R)} (2x) = \beta_{2x}, 
\nonumber
\\
\Psi^{(\pm,L)} (2x-1) &= \frac{e^{i \eta}}{\lambda_{\pm}} \beta_{2x}, \>\> \Psi^{(\pm,R)} (2x+1) = - \frac{\triangle e^{-i \eta}}{\lambda_{\pm}} \alpha_{2x} = \lambda_{\pm} e^{-i \eta} \alpha_{2x},
\label{huyumiHayaku2}
\end{align}
where $\alpha_{2x}, \> \beta_{2x} \in \CM$ with $\alpha_{2x} \beta_{2x} \not= 0$. In fact, we have 
\begin{align*}
U^{(s)} \Psi^{(\pm)} = \lambda_{\pm} \Psi^{(\pm)}.
\end{align*}
Then $\Psi^{(\pm)} \in {\cal W}^{(\lambda_{\pm})}$. Therefore
\begin{align}
(U^{(s)})^n \Psi^{(\pm)} = \lambda_{\pm}^n \Psi^{(\pm)}.
\label{huyumiHayaku3}
\end{align}

Let $\mu_n ^{(\Psi^{(\pm)})} = \phi ((U^{(s)})^n \Psi^{(\pm)})$ and  
\begin{align*}
\mu_n ^{(\Psi^{(\pm)})}
= {}^T \left[ \ldots, \mu_n^{(\Psi^{(\pm)})} (-2), \mu_n^{(\Psi^{(\pm)})} (-1), \mu_n^{(\Psi^{(\pm)})} (0), \mu_n^{(\Psi^{(\pm)})} (1), \mu_n^{(\Psi^{(\pm)})} (2), \ldots \right].
\end{align*}
From Eqs. \eqref{huyumiHayaku}, \eqref{huyumiHayaku2}, and \eqref{huyumiHayaku3}, we obtain  
\begin{align*}
\mu_n ^{(\Psi^{(\pm)})} = {}^T \left[ \ldots, |\alpha_{-2}|^2+|\beta_{-2}|^2, |\alpha_{-2}|^2+|\beta_0|^2, |\alpha_0|^2+|\beta_0|^2, |\alpha_0|^2+|\beta_2|^2, |\alpha_2|^2+|\beta_2|^2, \ldots \right].
\end{align*}
Therefore we see that for any $n \ge 0$, $\mu_n ^{(\Psi^{(\pm)})} = \mu_0 ^{(\Psi^{(\pm)})}$. So $\mu_0 ^{(\Psi^{(\pm)})}$ becomes a stationary measure, that is, $\mu_0 ^{(\Psi^{(\pm)})} \in {\cal M}_s (U)$. Moreover $\mu_n ^{(\Psi^{(\pm)})} (2x) = |\alpha_{2x}|^2+|\beta_{2x}|^2 \> (x \in \ZM)$. So in general, stationary measure $\mu_0^{(\Psi^{(\pm)})}$ is a non-uniform and non-exponential decay measure. Therefore we obtain
\begin{align*}
{\cal M}_s (U) \setminus \left( {\cal M}_{unif} \cup {\cal M}_{exp} \right) \not= \emptyset.
\end{align*}

\section{Case $b=0$ \label{bzero}}
In this case, we see that $U$ can be written as 
\[
U=
\begin{bmatrix}
e^{i \eta} & 0 \\
0 & \triangle e^{-i \eta} 
\end{bmatrix},
\]
where $\eta \in [0, 2 \pi)$ and $\triangle (= \det U) \in \mathbb C$ with $|\triangle|=1$. This section gives the following result which is stronger than Theorem \ref{biwako3}:
\begin{thm}
For any $U \in \mbox{\boldmath{U}} (2)$ with $b=0$, we have
\begin{align*}
{\cal M}_s (U)= {\cal M}_{unif} = {\cal M}_2 (U).
\end{align*}
\label{donasummer}
\end{thm}
From now on we will prove this theorem. By Eq. \eqref{sankeien}, we have
\begin{align}
\Psi^{L}_n (x) =  e^{i \eta n}  \Psi^{L}_0 (x+n), \quad 
\Psi^{R}_n (x) = \triangle^n e^{-i \eta n} \Psi^{R}_0 (x-n),
\label{sankeien2} 
\end{align}
for any $x \in \ZM$ and $n \ge 0$. Put 
\begin{align*}
a_x = |\Psi^{L}_0 (x)|^2, \quad b_x = |\Psi^{R}_0 (x)|^2.
\end{align*}
Remark that $\mu_0 (x) = a_x + b_x.$ Then we see that if $\mu_0 \in {\cal M}_s (U)$, then Eq. \eqref{sankeien2} gives 
\begin{align}
a_{x+n} + b_{x-n} = a_{x} + b_{x} \qquad (n \ge 1),
\label{sankeien3} 
\end{align}
for any $x \in \ZM$, since $\mu_n (x) = \mu_0 (x)$. When $n=1$, Eq. \eqref{sankeien3} becomes
\begin{align}
a_{x+1} + b_{x-1} = a_{x} + b_{x} \qquad (x \in \ZM).
\label{sankeien4} 
\end{align}
If we take $x \to x+1$ in Eq. \eqref{sankeien4}, then we have
\begin{align}
a_{x+2} +  b_{x} = a_{x+1} + b_{x+1} \qquad (x \in \ZM).
\label{sankeien5} 
\end{align}
Combining Eq. \eqref{sankeien4} with Eq. \eqref{sankeien5} gives
\begin{align}
a_{x+2} -  a_{x} = b_{x+1} - b_{x-1} \qquad (x \in \ZM).
\label{sankeien6} 
\end{align}
Put $c_1=a_0-b_{-1}$ and $c_2=a_1-b_0.$ Then by Eq. \eqref{sankeien6}, we have
\begin{align}
a_{2k}=b_{2k-1}+c_1,\quad a_{2k+1} = b_{2k}+c_2\qquad (k \in \ZM).
\label{sankeien7} 
\end{align}
Combining Eq. \eqref{sankeien7} and Eq. \eqref{sankeien3} with $x=2k$ and $n=2$ implies 
\begin{align}
a_{2k+2} -a_{2k+1} = a_{2k}-a_{2k-1}. 
\label{sankeien8}
\end{align}
Similarly, combining Eq. \eqref{sankeien7} and Eq. \eqref{sankeien3} with $x=2k+1$ and $n=2$ gives 
\begin{align}
a_{2k+3} -a_{2k+2} = a_{2k+1}-a_{2k}.
\label{sankeien9}
\end{align}
Let $A = a_1-a_0$ and $B = a_2-a_1.$ By Eqs. \eqref{sankeien8} and \eqref{sankeien9}, we get 
\begin{align}
 a_{2k} = (A+B)k+a_0,\quad a_{2k+1} = (A+B)k+ A + a_0 \qquad (k \in \ZM). \label{sankeien10}
\end{align}
If $A+B \neq 0$, then $a_x$ must be negative for some $x \in \ZM$. So we have $A+B=0$, and Eq. \eqref{sankeien10} becomes
\begin{align}
 a_{2k} =a_0,\quad a_{2k+1}=a_1 \qquad (k \in \ZM).\label{sankeien11}
\end{align}
In a similar fashion, we can obtain
\begin{align}
 b_{2k} =b_0,\quad b_{2k+1}=b_1 \qquad (k \in \ZM). \label{sankeien12}
\end{align}
By Eqs. \eqref{sankeien11} and \eqref{sankeien12},
\begin{align}
\mu_0(2k) =\mu_0(0),\quad \mu_0(2k+1)=\mu_0(1) \qquad (k \in \ZM). 
\label{sankeien13}
\end{align}
On the other hand, combining Eq. \eqref{sankeien3} with Eq. \eqref{sankeien12} gives
\begin{align}
\mu_0(0) = a_0+b_0 = a_1+b_{-1} = a_1+b_1=\mu_0(1). 
\label{sankeien14}
\end{align}
By Eqs. \eqref{sankeien13} and \eqref{sankeien14}, we have $\mu_0 (x) = \mu_0(0)= \mu_0(1)$ for any $x \in \ZM$, that is, $\mu_0$ is a uniform measure. So we see that $\mu_0=\mu_1=\mu_2$ implies that $\mu_0 \in {\cal M}_{unif}$. Then we obtain
\begin{align}
{\cal M}_2 (U) \subseteq {\cal M}_{unif}.
\label{lastdance1}
\end{align}
From Theorem \ref{biwako1} which will be shown in the next section, we have 
\begin{align}
{\cal M}_{unif} \subseteq {\cal M}_s (U).
\label{lastdance2}
\end{align}
On the other hand, by definition, we get
\begin{align}
{\cal M}_s (U) \subseteq {\cal M}_2 (U).
\label{lastdance3}
\end{align}
By Eqs. \eqref{lastdance1}, \eqref{lastdance2}, and \eqref{lastdance3}, we have
\begin{align*}
{\cal M}_2 (U) \subseteq {\cal M}_{unif} \subseteq {\cal M}_s (U) \subseteq {\cal M}_2 (U).
\end{align*}
Therefore we conclude that
\begin{align*}
{\cal M}_2 (U) = {\cal M}_{unif} = {\cal M}_s (U).
\end{align*}
So the proof is completed.
\par
\
\par
We should note that Theorem \ref{donasummer} implies 
\begin{align*}
{\cal M}_1 (U) \supseteq {\cal M}_2 (U) = {\cal M}_3 (U) = \cdots = {\cal M}_s (U) = {\cal M}_{unif}.
\end{align*}
Here we give two non-uniform and non-stationary examples which satisfy $\mu_0= \mu_1$ and $\mu_1 \not= \mu_2$. That is, 
\begin{align*}
{\cal M}_1 (U) \setminus {\cal M}_2 (U) \not= \emptyset. 
\end{align*}
Remark that for the Markov chain, if $\mu_0 = \mu_1$, then $\mu_0 = \mu_n$ for any $n \ge 2$, that is, 
\begin{align*}
{\cal M}_1^{Mc} = {\cal M}_2^{Mc}  = {\cal M}_3^{Mc}  = \cdots = {\cal M}^{Mc}_s,
\end{align*}
where ${\cal M}_n^{Mc}$ is the set of measures for the Markov chain which corresponds to ${\cal M}_n$.

The first one is an unbounded measure $\mu_0$ with $\mu_0 (x) = a_x + b_x$ as follows:
\begin{align*}
a_x = 
\left\{ 
\begin{array}{cc}
2x & (x \ge 1), \\
1 & (x =0), \\
-2x+1 & (x \le -1), 
\end{array}
\right.
\end{align*}
and
\begin{align*}
b_x = 
\left\{ 
\begin{array}{cc}
2x+3 & (x \ge 1), \\
3 & (x =0), \\
-2x & (x \le -1). 
\end{array}
\right.
\end{align*}
Then we have 
\begin{align*}
\mu_0 (x) = \mu_1 (x) = 
\left\{ 
\begin{array}{cc}
4x+3 & (x \ge 1), \\
4 & (x =0), \\
-4x+1 & (x \le -1).
\end{array}
\right.
\end{align*}
However we see $\mu_2 (0)=8$.

The second one is a bounded measure $\mu_0$ with $0 \le \mu_0 (x) = a_x + b_x \le 2 \> (x \in \ZM)$:  
\begin{align*}
a_{2x} &= a_{2x+1} = \frac{1}{2} + \frac{1}{2^2} + \cdots + \frac{1}{2^{x+1}} \quad (x \ge 0),
\\
a_{-2x} &= a_{-(2x-1)} = \frac{1}{2} - \frac{1}{2^2} - \cdots - \frac{1}{2^{x+1}} \quad (x \ge 1),
\\
b_{2x+1} &= b_{2x+2} = \frac{1}{2} + \frac{1}{2^2} + \cdots + \frac{1}{2^{x+2}} \quad (x \ge -1),
\\
b_{-(2x+1)} &= b_{-2x} = \frac{1}{2} - \frac{1}{2^2} - \cdots - \frac{1}{2^{x+1}} \quad (x \ge 1).
\end{align*}
Remark that $\mu_0 (x) < \mu_0 (x+1)$ for any $x \in \ZM$ with $\lim_{x \to - \infty} \mu_0 (x) = 0$ and $\lim_{x \to \infty} \mu_0 (x) = 2$.

\section{Proof of Theorem \ref{biwako1} \label{unitary}}
This section gives a new and simple proof of Theorem \ref{biwako1}, i.e., ${\cal M}_{unif} \subset {\cal M}_s (U)$ for any $U \in \mbox{\boldmath{U}} (2).$ First we consider the following initial state: for any $x \in \ZM$, 
\begin{align*}
\Psi_{0} (x) = \varphi = \begin{bmatrix} \alpha  \\ \beta  \end{bmatrix}\in \CM^2,
\end{align*}
where $||\varphi||^2 = |\alpha|^2+|\beta|^2>0$. Remark that $\Psi_{0} (x)$ does not depend on the position $x$. Then we have 
\begin{align*}
\Psi_{1} (x) = P \Psi_{0} (x+1) +  Q \Psi_{0}(x-1) = (P+Q) \varphi = U \varphi.
\end{align*}
Similarly, we have 
\begin{align*}
\Psi_{2} (x) = P \Psi_{1} (x+1) +  Q \Psi_{1}(x-1) = (P+Q) U \varphi = U^2 \varphi.
\end{align*}
Then we obatin
\begin{align*}
\Psi_{n} (x) = U^n \varphi
\end{align*}
for any $n =0,1,2, \ldots$ and $x \in \ZM$. Thus we have
\begin{align*}
\mu_n (x) &= 
|| \Psi_{n} (x) ||^2 =|| U^n \varphi ||^2 = || U^n \Psi_{0} (x) ||^2 = ||\Psi_{0} (x) ||^2 \\
&= \mu_0 (x) = ||\varphi ||^2 (= |\alpha|^2 + |\beta|^2),
\end{align*}
since $U$ is unitary. That is, this measure $\mu_0$ satisfies $\mu_0 = \mu_{u}^{(c)}$ with $c=||\varphi ||^2$ and 
\begin{align*}
\mu_n (x)= \mu_0 (x) \>\> (n \ge 1, \> x \in \ZM).
\end{align*}
Therefore the proof is completed. That is, we showed the following result: 
for any $U \in \mbox{\boldmath{U}} (2)$, we have
\begin{align*}
{\cal M}_{unif} \subseteq {\cal M}_s (U).
\end{align*}
We can easily generalize this theorem. For example, we consider an $N$-state QW on $\ZM$ determined by the $N \times N$ unitary matrix $U = [U(i,j)]_{1 \le i,j \le N}$. For $k=1,2, \ldots, N$, we put 
\begin{align*}
U_k (i,j) = U (i,j) \delta_{i,k}.
\end{align*}
Remark that $U$ is divided into $\{U_k : k=1,2, \ldots ,N \}$, i.e., 
\begin{align*}
U = \sum_{k=1}^N U_k. 
\end{align*}
For $N=2M+1$ with $M=1,2, \ldots$, $U_k$ corresponds to the weight of jump from $x$ to $x - M+ (k-1)$, where $k=1,2, \ldots, N(=2M+1)$. So the range of the jump is $\{x-M, x-M+1, \ldots, x+M-1, x+M\}$. For example, $M=1$ case is the three-state QW. Similarly, for $N=2M$ with $M=1,2, \ldots$, $U_k$ corresponds to the weight of jump from $x$ to $x - M + (k-1) \>\> (k=1,2, \ldots, M)$, and from $x$ to $x - M + k \>\> (k=M+1,M+2, \ldots, N=2M)$. So the range of the jump is $\{x-M, x-M+1, \ldots, -1, 1, \ldots , x+M-1, x+M\}$. Then $M=1$ case is the two-state QW considered here. In a general case also, we have the same argument as both of the $M=1$ case. Let $\Psi_0 = {}^T [\ldots, \varphi, \varphi, \varphi, \ldots]$, that is, $\Psi_0 (x)$ is always $\varphi={}^T\![\alpha_1, \alpha_2, \ldots, \alpha_N]$. Then we obtain 
\begin{align*}
\Psi_{1} (x) =  \sum_{k=1}^N U_k \varphi = U \varphi.
\end{align*}
In a similar way, we see
\begin{align*}
\Psi_{n} (x) = U^n \varphi.
\end{align*}
Therefore 
\begin{align*}
\mu_n (x)= || \Psi_{n} (x) ||^2 =|| U^n \varphi ||^2 = || U^n \Psi_{0} (x) ||^2 = ||\Psi_{0} (x) ||^2 = \mu_0 (x) (= ||\varphi||^2).
\end{align*}
Thus 
\begin{align*}
\mu^{(||\varphi||^2)}_{u} \in {\cal M}_s (U).
\end{align*}
Then it follows from this that for any $U \in \mbox{\boldmath{U}} (N)$, we see
\begin{align*}
{\cal M}_{unif} \subseteq {\cal M}_s (U).
\end{align*}

\section{Summary \label{sum}}
In this paper, we proved that
\par\noindent
(i) for any $U \in \mbox{\boldmath{U}} (2)$ with $abcd \not= 0$ or $a=0$, we have
\begin{align*}
{\cal M}_s (U) \setminus \left( {\cal M}_{unif} \cup {\cal M}_{exp} \right) \not= \emptyset,
\end{align*}
\par\noindent
(i) for any $U \in \mbox{\boldmath{U}} (2)$ with $b=0$, we have
\begin{align*}
{\cal M}_s (U) = {\cal M}_{unif}.
\end{align*}
One of the interesting future problems is to characterize ${\cal M}_s (U)$ for both $abcd \not= 0$ and $a=0$ cases, like $b=0$ case. Moreover, to consider ${\cal M}_s (U)$ for general $U \in \mbox{\boldmath{U}} (N)$ is a challenging problem.

\par
\
\par\noindent
{\bf Acknowledgments.} We would like to thank Kei Saito for useful discussions. This work was partially supported by the Grant-in-Aid for Scientific Research (C) of Japan Society for the Promotion of Science (Grant No.24540116).

\par
\
\par

\begin{small}
\bibliographystyle{jplain}

\begin{thebibliography}{99}




\bibitem{Kempe2003} 
Kempe, J.: 
Quantum random walks - an introductory overview. Contemporary Physics {\bf 44},  307--327 (2003) 


\bibitem{Kendon2007} 
Kendon, V.: 
Decoherence in quantum walks - a review. 
Math. Struct. in Comp. Sci. {\bf 17}, 1169--1220 (2007)


\bibitem{VAndraca2008} 
Venegas-Andraca, S. E.: 
Quantum Walks for Computer Scientists. Morgan and Claypool (2008)


\bibitem{Venegas2013} 
Venegas-Andraca, S. E.: 
Quantum walks: a comprehensive review.  
Quantum Inf. Process. {\bf 11}, 1015--1106 (2012)



\bibitem{Konno2008b} 
Konno, N.: 
Quantum Walks. In: Quantum Potential Theory, Franz, U., and Sch\"urmann, M., Eds., Lecture Notes in Mathematics: Vol. 1954, pp. 309--452, Springer-Verlag, Heidelberg (2008)



\bibitem{CanteroEtAl2013} 
Cantero, M. J., Gr\"unbaum, F. A., Moral, L., Vel\'azquez, L.: 
The CGMV method for quantum walks.   
Quantum Inf. Process. {\bf 11}, 1149--1192 (2012)



\bibitem{MW2013}
Manouchehri, K., Wang, J.:
Physical Implementation of Quantum Walks. 
Springer (2013) 



\bibitem{P2013}
Portugal, R.:
Quantum Walks and Search Algorithms. 
Springer (2013) 


\bibitem{AmbainisEtAl2001} 
Ambainis, A., Bach, E., Nayak, A., Vishwanath, A., Watrous, J.: 
One-dimensional quantum walks. In: Proceedings of the 33rd Annual ACM Symposium on Theory of Computing, pp. 37--49 (2001)



\bibitem{KLS2013}
Konno, N., {\L}uczak, T., Segawa, E.: 
Limit measures of inhomogeneous discrete-time quantum walks in one dimension. 
Quantum Inf. Process. {\bf 12}, 33--53 (2013)



\bibitem{EK2013}
Endo, T., Konno, N.: 
The stationary measure of a space-inhomogeneous quantum walk on the line. 
Yokohama Mathematical Journal (in press), arXiv:1309.3054 (2013)

 
\bibitem{WojcikEtAl2004} 
W\'ojcik, A., {\L}uczak, T., Kurzy\'nski, P., Grudka, A., Gdala, T., Bednarska-Bzdega, M.: 
Trapping a particle of a quantum walk on the line.  
Phys. Rev. A {\bf 85}, 012329 (2012)




\bibitem{EndoEtAl2014a}
Endo, T., Konno, N., Segawa, E., Takei, M.:
A one-dimensional Hadamard walk with one defect. 
Yokohama Mathematical Journal (in press), arXiv:1407.8103 (2014)



\bibitem{EndoEtAl2014b}
Endo, S., Endo, T., Konno, N., Segawa, E., Takei, M.:
Limit theorems of a two-phase quantum walk with one defect. 
arXiv:1409.8134 (2014)










\bibitem{Konno2014} 
Konno, N.: 
The uniform measure for discrete-time quantum walks in one dimension. 
Quantum Inf. Process. {\bf 13}, 1103--1125 (2014)





\end{thebibliography}

\end{small}

\end{document}